\documentstyle[12pt]{article}

\newcommand\beq{\begin{equation}}
\newcommand\eeq{\end{equation}}

\def\C{{\cal C}}

\begin{document}

\title{\vspace{-1cm}
The SU($N$) Wilson Loop Average \\ in 2 Dimensions\\
\vspace{.2cm}}{}{}

\author{
Esa Karjalainen\thanks{Karjalainen@pupgg.princeton.edu}\\
Joseph Henry Laboratories,\\ Department of Physics,\\
Princeton University,\\ Princeton, NJ 08544, USA}

\date{June 1993}

\maketitle
\vspace{-11.8cm}
\hfill PUPT-92/1360

\hfill hep-th/9306084
\vspace{10cm}

\begin{abstract}
We solve explicitly a closed, linear loop equation for the SU(2)
Wilson loop average on a two-dimensional plane and generalize the solution
to the case of the SU($N$) Wilson loop average with an arbitrary
closed contour. Futhermore, the flat space solution is generalized
to any two-dimensional manifold for the SU(2) Wilson loop average
and to any two-dimensional manifold of genus 0 for the SU($N$) Wilson
loop average. The SU($N$) Wilson loop average follows an area law
$W(\C) = \sum_r P_r^{'} e^{-\sum_i j_{ri}^2 S_i}$ where $j_{ri}^2$
is the quadratic Casimir operator for the window with area $S_i$.
Only certain combinations of the Casimir operators are allowed in the
sum over $i$. We give a physical interpretation of the constants
$P_r^{'}$ in the case of a non-self-intersecting composed path $\C$
and of the constraints determing in which combinations the Casimir
operators occur.
\end{abstract}
\newpage



\section{Introduction}

In a recent paper \cite{KarMig} it was shown that the loop equation
for the SU(2) Wilson loop average (WLA)
\beq
\label{defWLA}
W(\C_1,...,\C_k) := \langle \prod_{i=1}^k \frac{1}{N} {\rm Tr} P
e^{\oint_{\C_i} dx_{\mu} A_{\mu} } \rangle ,
\eeq
where $N=2$ and $P$ stands for path ordering along a closed contour
$\C_i$, is linear and closes for WLAs with the same number
of traces $k$. This raises hopes that the solution can be obtained.
In earlier papers it was proved that it is possible to solve the U($N$)
(or SU($N$)) loop equation in two dimensions by iteration \cite{Kaz}
and in the large $N$ limit the form of the solution was obtained showing
modified area law behavior with area dependent polynomials multiplying the
exponentials \cite{KazKos}. The WLAs in two dimensions can also be calculated
using an algorithm based on a non-Abelian version of Stokes theorem
\cite{Bra} or by expanding a lattice theory in the characters and taking
the continuum limit \cite{Mig75,Rus,Wit}. During the preparation of this
work we recieved a preprint \cite{GroTay} which
gives a general formula for the SU($N$) WLA on an arbitrary two dimensional
manifold in terms of maps of an open string world sheet onto the target
space. In this paper, we solve the closed, linear SU(2) loop equation in two
dimensional Euclidean space and generalize the solution to the case of the
SU($N$) (or U($N$)) WLA on a two-dimensional manifold of genus 0 and the
SU(2) (or U(2)) WLA on any two-dimensional manifold. We also give a physical
interpretation of the constants in the solution for a WLA with a
non-self-intersecting composed contour.
Because the WLA in the large distance limit follows the area law,
confinement takes place \cite{Wil}. On the other hand if the WLA would follow
the perimeter law as in four-dimensional
QED the interaction would decrease sufficiently
rapidly at large distances that there would be no confinement.
Because confinement takes place all the physical information in two
dimensional QCD can
be derived from the WLAs by proper integration over the loop space.

This paper is organized as follows. In section 2, we briefly review
the Migdal-Makeenko loop equation in any dimension and give the definitions
of the path and area derivative. In section 3, we solve the loop equation
for the SU(2) WLA on a two-dimensional plane. In section 4, the solution
is generalized to the case of SU($N$) WLA on a two-dimensional plane.
In section 5, we discuss the U($N$) WLA, consider the large $N$ limit and
generalize the flat space solution to any two-dimensional manifold for
the SU(2) (or U(2)) WLA and to any two-dimensional manifold of genus 0
for the SU($N$) (or U($N$)) WLA. Finally, we summarize our results and
discuss some interesting open questions.

\section{The loop equations}

For U($N$) ($S=0$) or SU($N$) ($S=1$) the Migdal-Makeenko loop equation
is linear (for finite $N$) and reads \cite{MakMig,Mig}
\beq
\partial_{\mu}(x) \frac{\delta}{\delta \sigma_{\mu \nu} (x)}
W(\C) =
g^2 \oint_{\C_{xx}} dy_{\nu} \delta^d (x-y) [N W(\C_{xy},\C_{yx})
- \frac{S}{N} W(\C)]
\eeq
where $\C_{xy}$ is a path from a point $x$ to a point $y$,
$\C = \C_{xy}\C_{yx}$ is a closed path and $g$ is the coupling
constant. An example of the paths $\C_{xy}$, $\C_{yx}$ and $\C$
with $x=y$ is shown in Fig. 1. The path derivative
$\partial_{\mu} (x)$ is defined as follows
\beq
\lim_{z \rightarrow y} (z_{\mu} - y_{\mu}) \partial_{\mu} (y)
F(\C_{xy}) := \lim_{z \rightarrow y} [F(\C_{xy}\C_{yz}) - F(\C_{xy})]
\eeq
for an open path $\C_{xy}$ and
\beq
\lim_{z \rightarrow y} (z_{\mu} -y_{\mu}) \partial_{\mu} (y)
F(\C_{yy}) := \lim_{z \rightarrow y} [F(\C_{zy} \C_{yy} \C_{yz}) -
F(\C_{yy})]
\eeq
for a closed path $\C_{yy}$, where $F$ is an arbitrary functional of
its argument. The area derivative $\frac{\delta}
{\delta \sigma_{\mu \nu} (x)}$ is defined as
\beq
\Delta \sigma_{\mu \nu}
\frac{\delta}{\delta \sigma_{\mu \nu} (x)} F(\C) :=
F(\C\C_{\sigma}) - F(\C)
\eeq
where $\Delta \sigma_{\mu \nu}$ is the oriented area of an infinitesimal loop
$\C_{\sigma}$.

For large $N$ the double trace WLA factorizes as
\beq
W(\C_{xy},\C_{yx}) = W(\C_{xy}) W(\C_{yx}) + {\cal O}(\frac{1}{N^2})
\eeq
which yields a closed but non-linear loop equation \cite{MakMig}.

For the SU(2) gauge field $A_\mu$ \cite{RovSmo}
\beq
\label{traceeqn}
TrPe^{\int_{\C_{xy}} dx_\mu A_\mu} TrPe^{\int_{\C_{yx}} dx_\mu A_\mu}
= TrPe^{\int_\C dx_\mu A_\mu} +
TrPe^{\int_{\C_{xy}\C_{yx}^{-1}} dx_\mu A_\mu}
\eeq
where the path $\C_{yx}^{-1}$ is the path $\C_{yx}$
travelled in the opposite direction. Thus we obtain the loop
equation \cite{KarMig}
\beq
\partial_\mu (x) \frac{\delta}{\delta \sigma_{\mu \nu} (x)} W(\C) =
g^2 \oint_{\C_{xx}} dy_\nu \delta^d (x-y)
[\frac{1}{2} W(\C) + W(\C_{xy}\C_{yx}^{-1})]
\eeq
which is both closed and linear.

\section{The SU(2) Wilson loop average}

In two dimensions the equation for SU(2) WLA can be written in the following
form at a point of a self-intersection (at other points the right hand
side vanishes) by integrating along a short path
from $x-\Delta$ to $x+\Delta$ intersecting the
loop at the self-intersection point $x=y$ \cite{KazKos}
\beq
\frac{\delta W(\C)}{| \delta \sigma_{\mu \nu}(x+\Delta)|} +
\frac{\delta W(\C)}{| \delta \sigma_{\mu \nu}(x-\Delta)|} =
g^2[\frac{1}{2} W(\C) + W(\C_{xy}\C_{yx}^{-1})]
\eeq
which can be further simplified to read
\beq
\label{loopeqn}
\frac{D}{DS_x} W_v(\C) :=
\Bigl( \frac{\delta}{\delta S_a} + \frac{\delta}{\delta S_b} -
\frac{\delta}{\delta S_c} - \frac{\delta}{\delta S_d} - \frac{1}{2}
\Bigr) W_v(\C) = W_{v-1}(\C_{xy}\C_{yx}^{-1})
\eeq
where the areas of the windows $S$ of the loop touching the point of
self-crossing are measured in units of $g^2$. The equation (\ref{loopeqn})
tells us the relation between the WLA $W_v(\C)$ with $v$ self-intersections
and the WLA
\linebreak
$W_{v-1}(\C_{xy}\C_{yx}^{-1})$ with $v-1$ self-intersections
where the self-intersection touching the areas $S_a$, $S_b$, $S_c$ and
$S_d$ has been broken so that the areas corresponding to the operators
multiplied by a minus sign are connected (Fig. 2).

By repeated use of the equation (\ref{loopeqn})
we can relate the WLA $W_v$ to
$v$ WLAs $W_{v-1}$ by breaking all the self-crossings and each of
the WLAs $W_{v-1}$ to $v-1$ WLAs $W_{v-2}$ etc. Thus we obtain
a tree of WLAs related to each other by partial differential equations
with $v!$ WLAs at the top of the tree.

In two dimensions the WLA $W(\C)$
depends on the loop $\C$ only through the areas of the windows formed
by the loop i.e.
\beq
\label{Sdepend}
W(\C) = W(S_a,S_b,...)
\eeq
which can be proved using diagram techniques (cf. (\ref{propagator}))
or by lattice theory
\cite{Kaz,KazKos}.
{}From the definition of the WLA (\ref{defWLA}) we obtain the boundary
conditions
\beq
\label{boundcond}
W(S_a,S_b,...) \biggr|_{S_a=0,S_b=0,...} = 1 .
\eeq
In addition, it can be seen (for example by an explicit calculation
in the axial gauge $A_1=0$) that the WLA does not depend on the
infinite area $S_0$ external to the loop. Thus
\beq
\frac{\delta W}{\delta S_0} = 0 .
\eeq

Now we shall prove by induction that the WLA has the following
form
\beq
\label{genform}
W_v(S_1,S_2,...) = \sum_r P_r^{'} e^{-\sum_i j_{ri}(j_{ri} + 1) S_i}
\eeq
where $P_r^{'}$s and $j_{ri}(j_{ri} + 1)$s are constants. (The suggestive form
of the latter will become clear shortly.)
In the axial gauge $A_1=0$ the propagator
\beq
\label{propagator}
G_{\mu\nu}(x-y) \propto |x_1 - y_1| \delta(x_2-y_2)
\eeq
and because $[A_2(x),A_2(x)] \equiv 0$ there are no gluon self-interactions
in two dimensions. Thus, for the non-nested WLA with one
self-intersection shown in Fig. 1,
\beq
W_1(S_a,S_b) = W_0(S_a+S_b) .
\eeq
The loop equation (\ref{loopeqn}) yields
\beq
\Bigl(- \frac{\delta}{\delta S_a} - \frac{\delta}{\delta S_b} - \frac{1}{2}
\Bigr)W_1(S_a,S_b) = W_0(S_a+S_b)
\eeq
which can be solved
\beq
W_0(S) = e^{-\frac{3}{4}S} .
\eeq
Thus $W_0(S)$ satisfies the assumption (\ref{genform}). Now we shall assume
that $W_{v-1}$ also satisfies the equation (\ref{genform}). The special
solution $W_v^s$ of the loop equation (\ref{loopeqn}) is now
\beq
W_v^s = \Bigl( \frac{D}{DS} \Bigr)^{-1}W_{v-1} = \frac{1}{C} W_{v-1}
\eeq
with $C \neq 0$ because vanishing of the constant $C$ would imply
that $W_{v-1} \equiv 0$ which is not possible according to the
boundary condition (\ref{boundcond}). The special solution $W_v^s$ has the
right form and so does the homogeneous solution obtained by
separating the variables. Thus we have proved equation (\ref{genform}).

Applying the operator $\frac{\delta}{\delta S_c} -
\frac{\delta}{\delta S_d}$ to equation (\ref{loopeqn}) makes the right
hand side vanish because $W_{v-1} = W_{v-1}(S_a,...S_b,S_c+S_d)$.
By operating on the tree of a WLA $W_v$ with these operators
it can be seen that $j_{ri} \in \{0,\frac{1}{2},1,...,\frac{Q}{2}\}$
subject to the condition that $j_{ri} = j_{ri'} \pm \frac{1}{2}$
for neighbouring windows $S$ and $S'$. $Q$ is
the minimum number of times the contour $\C$ has to be crossed in order
to reach the external area $S_0$ from the window $S_i$.
Thus the exponent is proportional to the time integral of the
one-dimensional string potential between the spatially separated
quark lines \cite{Mig75,Str}
\beq
V(t) \propto g^2 j(j+1) |x(t)| .
\eeq
The spin $j$ is obtained by the following angular momentum
addition rules. The SU(2) quarks transform in the fundamental
spin $s = \frac{1}{2}$ representation of
the gauge group. Thus $j = s = \frac{1}{2}$ for the areas $S$
surrounded by a single quark line, $j' = |j - s|,...,j + s = 0,1$
for the areas $S'$ surrounded by two quark lines etc.
Note that the limit $\lim_{S \rightarrow \infty} W \neq 0$ for
windows S surrounded by
pairs of quark lines, as expected. A quark and an adjacent
antiquark, represented by a pair of antiparallel lines, can
form a colorless ($j=0$) state i.e. a meson. Similarly, in the
case of the gauge group SU(2) two adjacent quarks, represented
by a pair of parallel lines, can form a colorless state i.e. a
baryon.

The constants $P_r^{'}$ can be solved from the boundary conditions
(\ref{boundcond})
for the WLAs in the tree of $W_v$ but this approach would require
us to draw the tree with of order $v!$ WLAs for every WLA $W_v$
we want to solve. Instead we can relate each term in the sum of
equation (\ref{genform}) to the term with the lowest possible exponents
of a nested WLA $W^{\rm nested}$ (Fig. 3) by breaking and adding
self-intersections. Next, we calculate the constant
$P_{\frac{1}{2} 1... \frac{v}{2}}^{\rm nested}$ multiplying the
exponential with the most negative exponent of a nested WLA.
We know that
\begin{eqnarray}
& W_v^{\rm nested}(S_1,S_2,...,S_{v+1}) = &
\nonumber \\
& \sum_{...j_Q = 0,\frac{1}{2},...,\frac{Q}{2};...{\rm such \, that} \,
j_{Q+1} = j_Q \pm \frac{1}{2} } P_{j_1 j_2 ... j_{v+1} }^{\rm nested}
e^{-\sum_{Q=1}^{v+1} j_Q (j_Q + 1) S_Q} . &
\end{eqnarray}
Also
\beq
W_{v-1}^{\rm nested} = \lim_{S_{v+1} \rightarrow 0} W_v^{\rm nested} .
\eeq
If $j_v = \frac{v}{2}$ then $j_{v+1} = \frac{v}{2} \pm \frac{1}{2}$ thus
\beq
P_{\frac{1}{2} 1... \frac{v}{2} }^{\rm nested} =
P_{\frac{1}{2} 1... \frac{v}{2} \frac{v+1}{2} }^{\rm nested} +
P_{\frac{1}{2} 1... \frac{v}{2} \frac{v-1}{2} }^{\rm nested} .
\eeq
On the other hand operating with $\frac{D}{DS}$ on the innermost
vertex of $W_v^{\rm nested}$ yields
\beq
P_{\frac{1}{2} 1... \frac{v}{2} }^{\rm nested} =
-(v+1) P_{\frac{1}{2} 1... \frac{v}{2} \frac{v-1}{2} }^{\rm nested}.
\eeq
Thus
\beq
P_{\frac{1}{2} 1... \frac{v+1}{2}}^{\rm nested} =
\frac{v+2}{v+1} P_{\frac{1}{2} 1... \frac{v}{2}}^{\rm nested}
\eeq
which yields
\beq
P_{\frac{1}{2} 1... \frac{v}{2} }^{\rm nested} =
\frac{v+1}{2} P_{\frac{1}{2} }^{\rm nested} = \frac{v+1}{2} .
\eeq

Now we can write the explicit solution of the loop equation for
the SU(2) WLA in two dimensions
\begin{eqnarray}
& W_v(\C) = W_v(S_{11},...,S_{QI},...) = &
\nonumber \\
& \sum_{j_0 = 0;... j_{QI} = 0,\frac{1}{2},...,\frac{Q}{2};...
{\rm such \, that} \, j_{Q+1,I^{'} } = j_{QI} \pm \frac{1}{2} }
\frac{2 {\rm max} j_{QI} + 1}{2}
\prod B(...,j_{QI},...) \times &
\nonumber \\
& \prod J(...,j_{QI},...) e^{-\sum_{Q,I} j_{QI} (j_{QI} + 1)
S_{QI} } &
\end{eqnarray}
where $I$ is a label which distinguishes between different windows with
the same value of $Q$ and the constraint
$j_{Q+1,I^{'} } = j_{QI} \pm \frac{1}{2}$ in the sum
applies for neighbouring windows $S_{QI}$ and $S_{Q+1,I^{'} }$.
We have related a generic term of the WLA to the term of the nested WLA
$W_{2 {\rm max} j_{QI}-1}^{\rm nested}(S_1,...,S_{2 {\rm max} j_{QI}})$
with the smallest possible values of $j$.
There is a breaking factor $B(...,j_{QI},...)$ for every
self-intersection one has to break to get the nested WLA from the
original WLA $W_v(\C)$. The total number of factors $B$ in a term
is $v - (2 {\rm max} j_{QI} - 1)$. From equation (\ref{loopeqn}) we see that
connecting windows $S_c$ and $S_d$ (with $j_c = j_d$)
touching the same vertex yields a breaking factor
\beq
B(j_a,j_b,j_c) = \frac{1}{-j_a (j_a + 1) - j_b (j_b + 1)
+ 2j_c (j_c + 1) - \frac{1}{2} } .
\eeq
In addition we have joining factors $J(...,j_{QI},...)$ for the
remaining sets of windows with the same value of $j$ which
cannot be connected by breaking vertices. There are $w-1$
joining factors for a set of $w$ windows. From equation (\ref{loopeqn})
it follows that connecting windows $S_c$ and $S_d$ separated by
a neighbouring area $S_a$ (with $j_a = j_c \pm \frac{1}{2} $)
yields a joining factor
\beq
J(j_a,j_c) = \frac{-2j_c (j_c + 1) +2j_a (j_a + 1) - \frac{1}{2} }
{-2j_a (j_a + 1) + 2j_c (j_c + 1) -\frac{1}{2} } .
\eeq
There can be more than one area separating the windows $S_c$ and
$S_d$ meaning only that there are other areas to be joined first.
Note that we do not have to keep track of
the orientation of the loop at the point where a vertex is to
be broken or added because we can always break and add vertices
in a particular order so that the windows with the same value of $j$
become connected.

To demonstrate the method we will write the answer for the WLA
shown in Fig. 4
\begin{eqnarray}
& W(S_1,S_{21},S_{22},S_3) = &
\nonumber \\
& \sum_{j_{21} = 0,1; j_{22} = 0,1;
j_3 = \frac{1}{2},\frac{3}{2}; j_3 = j_{22} \pm \frac{1}{2} }
C_{j_{21} j_{22} j_3 } \times &
\nonumber \\
& e^{-\frac{3}{4} S_1 -j_{21}(j_{21} +1)S_{21}
-j_{22}(j_{22} +1) S_{22} -j_3(j_3 +1) S_3 } &
\end{eqnarray}
where
\begin{eqnarray*}
& &
\!\!\!\!
C_{00 \frac{1}{2} } = \frac{2j_3 +1}{2} B(j_1,j_1,j_{21} \! = \! j_0)
B(j_1,j_1,j_{21} \! = \! j_{22}) \times
\\ & &
\;\;\;\;\;\;\;\;\;\;\! B(j_{21},j_{21},j_3 \! = \! j_1)
\biggr|_{j_{21}=j_{22}=j_0 \equiv 0; j_3 = j_1 \equiv \frac{1}{2} },
\\ & &
\!\!\!\!
C_{01 \frac{1}{2} } = \frac{2j_{22} +1}{2} B(j_1,j_1,j_{21} \! = \! j_0)
B(j_{21},j_{21},j_3 \! = \! j_1)
\biggr|_{j_{21}=j_0 \equiv 0; j_3= j_1 \equiv 0; j_{22} = J_{22} =1 },
\\ & &
\!\!\!\!
C_{10 \frac{1}{2} } = \frac{2j_{21} +1}{2} B(j_{21},j_{21},j_3 \! = \! j_1)
J(j_1,j_{22} \! = \! j_0) \times
\\ & &
\;\;\;\;\;\;\;\;\;\;\! B(j_{21},j_{22},j_1 \! = \! j_1)
\biggr|_{j_{22} = j_0 \equiv 0; j_3= j_1 \equiv \frac{1}{2};
j_{21}= J_{21}= 1 },
\\ & &
\!\!\!\!
C_{11 \frac{1}{2} } = \frac{2j_{21} +1}{2} B(j_{21},j_{21},j_3 \! = \! j_1)
B(j_1,j_1,j_{22} \! = \! j_{21})
\biggr|_{j_3= j_1 \equiv \frac{1}{2}; j_{22}= j_{21}= J_{21}= 1 },
\\ & &
\!\!\!\!
C_{10 \frac{3}{2} } = \frac{2j_3 +1}{2} J(j_1,j_{22} \! = \! j_0)
B(j_{21},j_{22},j_1 \! = \! j_1)
\biggr|_{j_{22}= j_0 \equiv 0; j_1 \equiv \frac{1}{2};
j_{21}= J_{21}= 1; j_3= J_3= \frac{3}{2} }
\end{eqnarray*}
and
\begin{eqnarray*}
& &
C_{11 \frac{3}{2} } = \frac{2j_3 +1}{2} B(j_1,j_1,j_{22} \! = \! j_{21})
\biggr|_{j_1 \equiv \frac{1}{2}; j_{22}= j_{21}= J_{21}= 1;
j_3= J_3= \frac{3}{2} }.
\end{eqnarray*}
The right orientations
in the terms having the joining factor $J$
can be maintained by first adding a new vertex
(which yields the numerator of $J$) then breaking the vertex
between the windows $S_{21}$ and $S_{22}$ and finally breaking the
new vertex (which yields the denominator of $J$).
The solution satisfies the boundary condition (\ref{boundcond}),
as it should. Namely
\beq
W(0,0,0,0) = \frac{1}{4} -\frac{3}{4} -\frac{1}{4} -\frac{1}{4}
+1 +1 = 1.
\eeq

To this point, we have discussed WLAs $W(\C)$ with only a single trace
($k=1$). Equation (\ref{traceeqn}) implies that a general
WLA $W(\C_1,...,\C_k)$,
for which the path is composed of any number of closed contours
$\C_i$, can be written in terms of the single loop WLAs i.e.
\beq
\label{composedWLA}
W(\C_1,\C_2,...,\C_k) = W(\C_1,\tilde{\C}_2,...,\tilde{\C}_k ) =
\frac{1}{2^{k-1} } \sum_{p_2,...,p_k = \pm 1}
W(\C_1 \tilde{\C}_2^{p_2}...\tilde{\C}_k^{p_k} )
\eeq
where the loops $\tilde{\C}_i$
are formed from the loops $\C_i$ so that the areas of the windows and the
self-crossings do not change but a smallest possible number of
self-touchings are added to make the composite contour connected
(cf. Fig. 5a and 5b). The equality of the original WLA and the WLA with the
modified path follows from equation (\ref{Sdepend}). Equation
(\ref{composedWLA}) implies that
\beq
W(\C_1^{p_1},\C_2^{p_2},...,\C_k^{p_k}) = W(\C_1,\C_2,...,\C_k)
\eeq
for any combination of directions $p_i = \pm 1$. Later it will become clear
that the reason why the relative orientation of loops makes no
difference is that $N-1=1$ for SU(2). Furthermore,
it can be seen from equations (\ref{Sdepend}) and (\ref{propagator})
that
\beq
W(\C_1,...,\C_i,\C_{i+1},...,\C_k) = W(\C_1,...,\C_i)
W(\C_{i+1},...,\C_k)
\eeq
for two (composite) contours $\C_1,...,\C_i$ and $\C_{i+1},...,\C_k$
windows of which do not overlap. The value of $j$
in the exponents for a general WLA follows the same angular momentum
addition rules as for the single loop WLA.

\section{The SU($N$) Wilson loop average}

In order to be able to generalize the result to SU($N$)
we rewrite the answer for a WLA with a non-self-intersecting
composed path. The WLA reads
\begin{eqnarray}
& W(S_{11},...,S_{QI},...) = &
\nonumber \\
& \sum_{j_0 = 0;... j_{QI} = 0,\frac{1}{2},...,\frac{Q}{2};...
{\rm such \, that} \, j_{Q+1,I^{'} } = j_{QI} \pm
\frac{1}{2} } P_{j_{11}...j_{QI}...} e^{-\sum_{Q,I} j_{QI}(j_{QI}+1)
S_{QI} } . &
\end{eqnarray}
The constant $P_{j_{11}...j_{QI}...}$ is the probability of the
combination $j_{11},...,j_{QI},...$ for the spins $j$ and it can be
calculated as follows. The multiplicity of a spin $j$ state is
$D(j) = 2j+1$. If the spin for the window $S_{QI}$ is $j_{QI}$
then the probability, that the spin $j_{Q+1,I^{'} }$ for the neighbouring
window $S_{Q+1,I^{'} }$ equals $j_{QI} \pm \frac{1}{2}$, is
\beq
P(j_{Q+1,I^{'} } = j_{QI} \pm \frac{1}{2}) =
\frac{D(j_{QI} \pm \frac{1}{2}) }{D(j_{QI} - \frac{1}{2}) +
D(j_{QI} + \frac{1}{2}) } = \frac{D(j_{Q+1,I^{'} }) }{2D(j_{QI}) } .
\eeq
Note that the above equation is also valid for $j_{QI} = 0$ because
$D(-\frac{1}{2}) = 0$. Thus we obtain
\beq
P_{j_{11}...j_{QI}...} = \frac{ \prod_{j_{QI} \in
\{j^{'}|S^{'} {\rm simply \, connected} \} } D(j) }
{2^L \prod_{j_{QI} \in \{ j^{'}|S^{'} {\rm not \, simply \, connected} \} }
[D(j)]^{w_{QI}-1} }
\eeq
where $L$ is the number of loops and $w_{QI}$ is the number of windows
$S_{Q+1,I^{'} }$ surrounded by the window $S_{QI}$.

The solution for a WLA with a non-self-intersecting composed path can
be generalized to the case of the gauge group SU($N$) and it reads
\begin{eqnarray}
\label{nsiWLA}
& W(...,S_{QRI},...) = &
\nonumber \\
& \sum_{...,[n_1,n_2,...]_{QRI},...} P_{...[n_1,n_2,...]_{QRI}...}
e^{-\sum_{Q,R,I} j^2 [n_1,n_2,...]_{QRI} S_{QRI} } . &
\end{eqnarray}
The quadratic Casimir operator $j^2 [n_1,n_2,...]$, where $n_i$ is the
nAumber of boxes in the $i$th column of the Young tableau for the
representation $[n_1,n_2,...]$, is given by the well known formula
\beq
j^2 [n_1,n_2,...] = \frac{1}{2} [ N \sum_i n_i - \sum_i n_i (n_i +1 -2i)
- \frac{S}{N} ( \sum_i n_i )^2 ]
\eeq
where $S=1$ for SU($N$) (and $S=0$ for U($N$)).
For SU(3) (with $n_i =2$ if $1 \leq i \leq r$ and $n_i =1$ if
$r < i \leq r+q$) the expression for the quadratic Casimir
operator simplifies to
\beq
j^2 [n_1,n_2,...] = \frac{1}{3}(q^2 + qr + r^2) +q +r .
\eeq
Let us connect the areas $S_{00}$ and $S_{QRI}$ with a directed open path
$L_{QRI}$ from the external infinite area $S_{00}$ to the window $S_{QRI}$ so
that the contour crosses the path as few times as possible. Then Q (R) is the
number of times the contour crosses the path from left to right (from right
to left). Examples of the paths $L_{QRI}$ and the indices $Q$ and $R$ are
shown in Fig. 6.
The following contraints apply to the sum over representations. The first
constraint on the representation $[n_1,n_2,...]_{QRI}$ is that it must be
one of the elements of the direct sum obtained by expanding
$([1] \otimes)^Q ([N-1] \otimes)^R$ i.e.
\beq
\label{flatconstr}
[n_1,n_2,...]_{QRI} \in \{[m_1,m_2,...] \,|\,
([1] \otimes)^Q ([N-1] \otimes)^R
= ... \oplus [m_1,m_2,...] \oplus ... \} .
\eeq
In addition, the representations $[n_1,n_2,...]_{Q+1,RI^{'} }$ and
$[l_1,l_2,...]_{QRI}$ for neighbouring windows $S_{Q+1,RI^{'} }$
and $S_{QRI}$ are related. $[n_1,n_2,...]_{Q+1,RI^{'} }$ is an
element of the direct sum obtained by expanding
$[l_1,l_2,...]_{QRI} \otimes [1]$ i.e.
\begin{eqnarray}
& [n_1,n_2,...]_{Q+1,RI^{'} } \in &
\nonumber \\
& {\cal A} := \{[m_1,m_2,...] \,|\,
[l_1,l_2,...]_{QRI} \otimes [1] = ...\oplus [m_1,m_2,...] \oplus ... \} . &
\end{eqnarray}
Finally, the representations $[n_1,n_2,...]_{Q,R+1,I^{'} }$
and $[l_1,l_2,...]_{QRI}$ for neighbouring windows $S_{Q,R+1,I^{'} }$
and $S_{QRI}$ are related. $[n_1,n_2,...]_{Q,R+1,I^{'} }$ is an
element of the direct sum obtained by expanding
$[l_1,l_2,...]_{QRI} \otimes [N-1]$ i.e.
\begin{eqnarray}
& [n_1,n_2,...]_{Q,R+1,I^{'} } \in &
\nonumber \\
& {\cal B} := \{[m_1,m_2,...]|
[l_1,l_2,...]_{QRI} \otimes [N-1] = ... \oplus [m_1,m_2,...] \oplus ... \} . &
\end{eqnarray}
As before a quark and an adjacent antiquark can form a colorless
($[N] = [0]$) state, i.e. a meson (because $[N-1] \otimes [1] = [N]
\oplus ...$), but in the case of SU($N$) gauge group $N$ adjacent quarks are
needed to form a colorless state, i.e. a baryon, as expected.
If the representation for the window $S_{QRI}$ is $[m_1,m_2,...]_{QRI}$
then the probability, that the representation is $[n_1,n_2,...]_{Q+1,RI^{'} }$
for the neighbouring window $S_{Q+1,RI^{'} }$, reads
\begin{eqnarray}
& P([n_1,n_2,...]_{Q+1,RI^{'} } | [m_1,m_2,...]_{QRI} ) = &
\nonumber \\
& \frac{D([n_1,n_2,...]_{Q+1,RI^{'} }) }
{\sum_{[l_1,l_2,...]_{Q+1,RI^{'} } \in {\cal A} }
D([l_1,l_2,...]_{Q+1,RI^{'} }) }
= \frac{D([n_1,n_2,...]_{Q+1,RI^{'} }) }
{N D([m_1,m_2,...]_{QRI}) } &
\end{eqnarray}
because $D([1]) = N$. The dimension $D$ of a representation $[n_1,n_2,...]$
is given by Weyl's formula
\beq
D([n_1,n_2,...]) = \prod_{1 \leq i < k \leq N} \frac{r_i - r_k +k -i}{k-i}
\eeq
where $r_i$ is the number of boxes in the $i$th row of the Young tableau.
For SU(3) the expression simplifies to
\beq
D([n_1,n_2,...]) = \frac{1}{2} (q+1)(q+r+2)(r+1)
\eeq
Similarly,
\begin{eqnarray}
& P([n_1,n_2,...]_{Q,R+1,I^{'} }| [m_1,m_2,...]_{QRI}) = &
\nonumber \\
& \frac{D([n_1,n_2,...]_{Q,R+1,I^{'} }) }
{\sum_{[l_1,l_2,...]_{Q,R+1,I^{'}} \in {\cal B} }
D([l_1,l_2,...]_{Q,R+1,I^{'}}) }
= \frac{D([n_1,n_2,...]_{Q,R+1,I^{'} }) }
{N D([m_1,m_2,...]_{QRI}) } &
\end{eqnarray}
because $D([N-1]) = N$. Thus we obtain
\begin{eqnarray}
\label{nsiP}
& P_{...[n_1,n_2,...]_{QRI}...} = &
\nonumber \\
& \frac{\prod_{[n_1,n_2,...]_{QRI} \in \{[m_1,m_2,...]_{QRI}| S_{QRI} \,
{\rm simply \, connected} \} } D([n_1,n_2,...]_{QRI}) }
{N^L \prod_{[n_1,n_2,...]_{QRI} \in \{[m_1,m_2,...]_{QRI}| S_{QRI} \,
{\rm not \, simply \, connected} \} }
[D([n_1,n_2,...]_{QRI} )]^{w_{QRI} -1} } &
\end{eqnarray}
where $w_{QRI}$ is the number of windows $S_{Q+1,RI^{'} }$ and
$S_{Q,R+1,I^{'} }$ surrounded by the window $S_{QRI}$.

For a WLA with an arbitrary path $\C$ we find
\begin{eqnarray}
& W_v (\C) = W_v (...,S_{QRI},...) = &
\nonumber \\
& \sum_{...,[n_1, n_2,...]_{QRI},... } P_{...[n_1,n_2,...]_{QRI}... }^{'}
e^{-\sum_{Q,R,I} j^2 [n_1,n_2,...]_{QRI} S_{QRI} } &
\end{eqnarray}
with the same constraints for the sum over representations as before.
Recall that the constants $P^{'}$ can be interpreted as probabilities ($P$)
only when the loop is non-self-intersecting ($v=0$).
The expression for the constants $P^{'}$ can be easily derived for the case
in which $S_a \equiv S_b$.
In this case we can break every vertex
using the Migdal-Makeenko loop equations which yield
\beq
\label{Pcomma}
P^{'}_{...[n_1,n_2,...]_{QRI}...} = P_{...[n_1,n_2,...]_{QRI}...}
\prod_{i=1}^{v} N^{{\epsilon}_i} B_{+} (...,[n_1,n_2,...]_{Q_i R_i I_i},...)
\eeq
where ${\epsilon}_i = 1$ (${\epsilon}_i = -1$) if the number of loops
increases (decreases) when breaking a vertex and
the general breaking factor $B_{+}$ reads ($j^2_a = j^2_b$)
\begin{eqnarray}
& B_{+} ([l_1,l_2,...]_c,[m_1,m_2,...]_d,[n_1,n_2,...]_a) = &
\nonumber \\
& \frac{1}{-2j^2[n_1,n_2,...]_a +j^2[l_1,l_2,...]_c +j^2[m_1,m_2,...]_d +
\frac{S}{N} } . &
\end{eqnarray}
Note that the general SU($N$) breaking operation connects the areas
$S_a$ and $S_b$ in Fig. 2a and it does not change the orientation of
any part of the loop \cite{Kaz}. For a general self-intersecting loop,
in addition to breaking vertices, we need to take the limit
$S_{QRI} \rightarrow 0$ for various areas $S_{QRI}$ to relate a term of the
WLA $W$ to a term of a WLA with a non-self-intersecting path. The limit
$S_b \rightarrow 0$ gives
\beq
\sum_{[m_1,m_2,...]_b} P^{'}_{...[l_1,l_2,...]_a [m_1,m_2,...]_b
[n_1,n_2,...]_c...} = P^{'}_{...[l_1,l_2,...]_a \bullet [n_1,n_2,...]_c...}
\eeq
where the sum is over all the allowed representations for the Casimir
operator $j^2[m_1,m_2,...]_b$ when all the other representations are kept
fixed. The bullet means that $P^{'}_{...[l_1,l_2,...]_a \bullet
[n_1,n_2,...]_c...}$ is calculated for $\lim_{S_b \rightarrow 0} W$.

To demonstrate how the method works we determine the constant
\linebreak
$P^{'}_{...\alpha \beta_1 \gamma \delta \epsilon \phi \varphi_1}$ for
the WLA $W(\C) = W(...,S_{60},S_{701},S_{702},S_{703},S_{704},S_{80},S_{90})$
where $\alpha = [3,2,1]_{60}$, $\beta_1 = [3,2,2]_{701}$,
$\gamma = [3,3,1]_{702}$, $\delta = [3,2,2]_{703}$, $\epsilon =[3,3,1]_{704}$,
$\phi = [3,3,2]_{80}$ and $\varphi_1 = [3,3,2,1]_{90}$.
The loop $\C$ is shown in Fig. 7a. If we keep the representation $\phi$
fixed the allowed representations for $j^2_{90}$ are $\varphi_1 = [3,3,2,1]$,
$\varphi_2 = [3,3,3]$ and $\varphi_3 = [4,3,2]$. Because the border of the
window $S_{90}$ is a non-self-intersecting loop we obtain
\beq
\label{split}
P^{'}_{...\alpha \beta_1 \gamma \delta \epsilon \phi \varphi_1} =
\frac{D(\varphi_1)}{\sum_{i=1}^{3} D(\varphi_i) }
P^{'}_{...\alpha \beta_1 \gamma \delta \epsilon \phi \bullet } .
\eeq
If we keep the representations $\alpha$ and $\phi$ fixed the allowed
representations for $j^2_{701}$ are $\beta_1 = [3,2,2]$ and
$\beta_2 = [3,3,1]$ which yields
\beq
P^{'}_{...\alpha \beta_1 \gamma \delta \epsilon \phi \bullet} =
P^{'}_{...\alpha \bullet \gamma \delta \epsilon \phi \bullet} -
P^{'}_{...\alpha \beta_2 \gamma \delta \epsilon \phi \bullet} .
\eeq
Notice that $\gamma$ and $\delta$ fix the value of $\phi$. Thus
\beq
P^{'}_{...\alpha \bullet \gamma \delta \epsilon \phi \bullet} =
P_{...\alpha \bullet \bullet \delta \epsilon \bullet \bullet} =
\frac{D(\delta)D(\epsilon) }{N^8 D(\alpha) }
\eeq
where the index $\gamma$ has also disappeared because areas $S_{702}$ and
$S_{704}$ are joined in the limit $S_{80} \rightarrow 0$. Because
$\beta_2 = \gamma = \epsilon$, and $\gamma$ and $\delta$ fix the value of
$\phi$
\beq
P^{'}_{...\alpha \beta_2 \gamma \delta \epsilon \phi \bullet} =
[B_{+} (\gamma,\alpha,\phi)]^2
P_{...\alpha \bullet \bullet \delta \epsilon \bullet \bullet} .
\eeq
We have written the constant $P^{'}_{...\alpha \beta_1 \gamma \delta
\epsilon \phi \varphi_1}$ in terms of the dimensions $D$ of the
representations , the breaking factors $B_{+}$ and the probability
$P_{...\alpha \bullet \bullet \delta \epsilon \bullet \bullet}$
for a WLA with a non-self-intersecting contour shown in Fig. 7b.

Notice that first we have to calculate the probabilities
for the non-self-intersecting parts surrounded by a self-intersecting part
to be able to contract the non-self-intersecting parts (cf. equation
(\ref{split}) of the example above).
Also note that the representation $[n_1,n_2,...]_{QR}$ is limited
to one possibility if the representations $[l_1,l_2,...]_{Q \pm 1,RI}$ and
$[m_1,m_2,...]_{Q \pm 1,RI^{'} }$ (or $[l_1,l_2,...]_{Q,R \pm 1,I}$ and
\linebreak
$[m_1,m_2,...]_{Q,R \pm 1,I^{'} }$), for windows $S_{Q \pm 1,RI}$ and
$S_{Q \pm 1,RI^{'} }$ (or $S_{Q,R \pm 1,I}$ and
\linebreak
$S_{Q,R \pm 1,I^{'} }$) adjacent to the window $S_{QR}$,
are fixed and not the same. Namely for $[l_1,l_2,...]_{Q+1,RI} \neq
[m_1,m_2,...]_{Q+1,RI^{'} }$ (or $[l_1,l_2,...]_{Q,R-1,I} \neq$
\linebreak
$[m_1,m_2,...]_{Q,R-1,I^{'} }$)
\beq
[n_1,n_2,...]_{QR} = [{\rm min}(l_1,m_1), {\rm min}(l_2,m_2),...]
\eeq
if $\sum_i l_i = \sum_i m_i$ and
\beq
[n_1,n_2,...]_{QR} = [{\rm min}(N,m_1), {\rm min}(l_1,m_2),
{\rm min}(l_2,m_3),...]
\eeq
if $N + \sum_i l_i = \sum_i m_i$. For $[l_1,l_2,...]_{Q-1,RI} \neq
[m_1,m_2,...]_{Q-1,RI^{'} }$ (or \linebreak $[l_1,l_2,...]_{Q,R+1,I} \neq
[m_1,m_2,...]_{Q,R+1,I^{'} }$)
\beq
[n_1,n_2,...]_{QR} = [{\rm max}(l_1,m_1), {\rm max}(l_2,m_2),...]
\eeq
if $\sum_i l_i = \sum_i m_i$ and
\beq
[n_1,n_2,...]_{QR} = [{\rm max}(N,m_1), {\rm max}(l_1,m_2),
{\rm max}(l_2,m_3),...]
\eeq
if $N + \sum_i l_i = \sum_i m_i$. There are at most two possible
representations \linebreak $[n_1,n_2,...]_{QR}$ if the representations
$[l_1,l_2,...]_{Q-1,R}$ (or $[l_1,l_2,...]_{Q,R+1}$) and \linebreak
$[m_1,m_2,...]_{Q+1,R}$ (or $[m_1,m_2,...]_{Q,R-1}$) for windows
adjacent to the window $S_{QR}$ are fixed because there are at most two
different orders in which the two boxes can be added to make the Young
tableau of the representation $[m_1,m_2,...]_{Q+1,R}$ from the one for
$[l_1,l_2,...]_{Q-1,R}$. Thus we can break vertices to join windows with
the same representation, change representations so that windows can be
joined or take limits $S_{QRI} \rightarrow 0$ until there are no vertices.

\section{Extensions to U($N$), the large $N$ limit and curved manifolds}

Thus, we have shown that all SU($N$) (and in particular SU(3)) Wilson
loop averages obey the area law.
Therefore, two-dimensional QCD exhibits confinement.
In two dimensions even QED has confinement
because in one spatial dimension the electromagnetic field cannot
spread out and thus the field strength is constant. In four-dimensional
space-time the electromagnetic field decreases sufficiently at large
distances that there is no confinement but the gluon field strength
is conjectured to remain constant because gluon self-interactions
keep quarks confined. The well known expression for the U(1)
WLA showing the area law behaviour can be derived to all orders in
perturbation theory using the axial gauge $A_1 = 0$. It reads
\beq
W^{ {\rm U} (1)}(\C) = W^{ {\rm U} (1)}(...,S_{QRI},...) =
e^{-\frac{1}{2} \sum_{Q,R,I} (Q-R)^2 S_{QRI} }
\eeq
where $Q-R$ is the winding number of the loop around the
window $S_i$. The winding number can only have a single value
for each window of a loop which reflects the abelian character
of the gauge group U(1). Thus the U(1) WLA has only one term
for any loop (the SU($N$) WLA has a term corresponding to
each possible combination of the representations of the non-abelian
gauge field). An electron and an adjacent positron, represented by
antiparallel lines, form a chargeless ($Q=R$) state i.e.
a positronium atom. The U($N$) WLA in two dimensions is a product of the U(1)
and the SU($N$) WLAs because the U(1) and the SU($N$) parts of the
action decouple. Thus in general ($S=0$ for U($N$) and $S=1$ for SU($N$))
\begin{eqnarray}
\label{UWLA}
& W(\C) = [W^{ {\rm U}(1) }(\C)]^{\frac{1-S}{N} } W^{ {\rm SU}(N) }(\C) = &
\nonumber \\
& \sum_{...,[n_1,n_2,...]_{QRI},...} P^{'}_{...[n_1,n_2,...]_{QRI}...}
e^{-\sum_{Q,R,I} \{ \frac{1-S}{2N} (Q-R)^2 + j^2_1[n_1,n_2,...]_{QRI} \}
S_{QRI} } \; &
\end{eqnarray}
where $j^2_1$ is the SU($N$) Casimir operator.

In the large $N$ limit (with fixed $g^2 N$) the WLAs follow a modified
area law with area dependent ``constants'' $P^{'}$ \cite{Kaz}.
The reason for the modified behaviour can be easily seen by considering
the SU($N$) WLA for Fig. 5(c) \cite{Bra}. It reads
\beq
W(\C_1 \tilde{\C}_2) = W(S_1,S_2) = e^{-\frac{N^2 -1}{2N} g^2 S_1
-\frac{2N^2 -4}{2N} g^2 S_2 } [\frac{N+1}{2} e^{-g^2 S_2} -\frac{N-1}{2}
e^{g^2 S_2} ] .
\eeq
In the large $N$ limit it reduces to
\beq
\lim_{N \rightarrow \infty } W(S_1,S_2) =
e^{-\frac{N}{2} g^2 S_1 -N g^2 S_2 } (1 -N g^2 S_2)
\eeq
where the second term arises because $P^{'} \propto N^E$
for constant $E \geq 1$.
It can be shown by induction (by adding loops) that for a WLA with a
non-self-intersecting contour $P \propto N^{-2E}$ where $E \geq 0$.
Thus the WLAs with a non-self-intersecting contour do not have area
dependent ``constants'' in the large $N$ limit. But the breaking factor
$B_{+}$ increases the exponent $E$ by one if its denominator is proportional
to $N^0$. Thus modified area dependence is possible for WLAs with a
self-intersecting path.

Finally, we will consider the WLAs on an arbitrary two-dimensional manifold
of genus $G$ and generalize the solution in the case of
non-self-intersecting loops found in \cite{Rus} to the case of
arbitrary loops. A WLA on a non-orientable manifold coincides with a WLA on a
non-compact manifold. Thus we can concentrate only on orientable
manifolds \cite{Rus}. A WLA with a non-self-intersecting composed path has the
same form as the WLA in equation (\ref{UWLA}) except that the constant $P$ is
replaced by a more general constant $P^G$ and the constraint
(\ref{flatconstr}) does not apply because there is no external infinite area.
We can assign the values $Q=R=0$ to an arbitrary window to calculate the
indices $Q$ and $R$ for the other windows. The constant $P^G$ is given by
\cite{Rus}
\beq
P_{...[n_1,n_2,...]_{QRI}...}^G = \frac{ \prod_{Q,R,I}
[D([n_1,n_2,...]_{QRI})]^{2-H_{QRI} } }{N^L Z_G}
\eeq
where $H_{QRI}$ is the number of disconnected loops that form the border
of $S_{QRI}$ and the partition function $Z_G$ reads
\beq
Z_G = \sum_{[n_1,n_2,...]} [D([n_1,n_2,...])]^{2-2G}
e^{-j^2_S [n_1,n_2,...] \sum_{Q,R,I} S_{QRI} } .
\eeq
The partition function depends on the total area of the manifold and the
sum is over all the irreducible representations of SU($N$) (or U($N$)).
In the limit $S_{00} \rightarrow \infty$ the only term that survives
is the one with the one-dimensional trivial representation
($j^2 [0]_{00} = 0$) giving the constraint (\ref{flatconstr}) and making
$Z_G = 1$. In this limit on a manifold with no handles the WLA reduces to
the one given by equations (\ref{nsiWLA}) and (\ref{nsiP}).

For a SU(2) (or U(2)) WLA with a self-intersecting path the constant
$P^{' \, G}$ can be calculated by breaking and adding vertices, as on the flat
manifold, because the breaking operation is local and the WLA depends on the
path only through the areas of the windows (which can be seen by using the
technique in \cite{Mig75,Rus}). Note that
\beq
P_{\frac{1}{2} 1...\frac{v}{2} }^{{\rm nested} \, G} =
2^{v-1} P^G_{\frac{1}{2} 1...\frac{v}{2} } =
\frac{\prod_{Q=1}^{v} (Q+1)^{2-H_Q} }{2Z_G}
\eeq
which reduces to $\frac{v+1}{2}$ on Euclidean space.
The first equality above follows from equation (\ref{Pcomma}).
A general constant $P^{' \, G}_{...j_{QI}...}$ reads
\beq
P^{' \, G}_{...j_{QI}...} =
P^{{\rm nested} \, G}_{\frac{1}{2} 1...\frac{v}{2} } \prod B(...,j_{QI},...)
\prod J(...,j_{QI},...) .
\eeq
As before there is a breaking factor $B$ for every self-intersection one has
to break (and a joining factor $J$ for the remaining sets of windows with
the same value of $j$ which cannot be connected by breaking vertices)
to get the nested WLA with the constant
$P^{{\rm nested} \, G}_{\frac{1}{2} 1...\frac{v}{2} }$ from the original
WLA with the constant $P^{' \, G}_{...j_{QI}...}$. We chose to use $B$ and
$J$ (which is a product of $B(...,j_{QI},...)$ and
$[B(...,j_{Q^{'} I^{'} },...)]^{-1}$) rather than the two different breaking
factors $B$ and $B_{+}$
to relate a generic term of the WLA to a term of a WLA with a
non-self-intersecting contour so that we did not have to keep track
of the orientation of the contour.

The constant $P^{' \, G}$ for a SU($N$) (or U($N$)) WLA can be calculated
(by breaking vertices and by letting $S_{QRI} \rightarrow 0$ for various
windows $S_{QRI}$) only when the boundary of the window $S_{QRI}$ can be
shrunk to an empty set of points because only then can the vertices on the
boundary be eliminated. This is always possible on a manifold with no
handles ($G=0$) and we can write $P^{' \, G}$ in terms of the dimensions
$D$ of the representations, the breaking factors $B_{+}$ and the
probabilities $P^{G}$. Note that also on a compact manifold
$P^{0}_{...[n_1,n_2,...]_{QRI}...}$ can be interpreted as the relative
probability of the combination $...[n_1,n_2,...]_{QRI}...$ for the
representations.

\section{Conclusions}

We have solved explicitly the closed, linear loop equation for the SU(2)
WLAs on a two-dimensional plane and generalized the solution to case of
the SU($N$) (or U($N$)) WLA on a two-dimensional manifold of genus 0 and
the SU(2) (or U(2)) WLA on any two-dimensional manifold. The WLA
follows an area law $W(\C) = \sum_r P^{'}_r e^{-\sum_i C_{ri} S_i }$
where $C_{ri}$ is the quadratic Casimir operator for SU($N$) (plus a U(1)
term proportional to the winding number squared of the loop around the
window $S_i$ for U($N$)). Only certain combinations of the Casimir operators
are allowed in the sum over $i$. Namely, the representations of the Casimir
operators differ by one box (or $N-1$ boxes) in the Young tableau for
neighbouring windows. This means that $N$ quarks or a quark and an antiquark
can form a particle i.e. a baryon or a meson.
In the case of a non-self-intersecting composed path on a manifold with no
handles the constant $P_r$ can be interpreted as the probability
of the combination $r$ for the representations.

It would be interesting to calculate the meson spectrum in two dimensions
for SU(3) or SU(2) QCD and to compare it to the large $N$ spectrum \cite{tHo}.
This could be done along the lines of \cite{GroTay,Str} in a single sector
of the theory. The modified area law for the large $N$ WLAs suggests that
there might be important differences between the case with infinite $N$ and
the one with finite $N$. On the other hand, the WLAs with the modified area
law behaviour are not needed to calculate the large $N$ meson spectrum to
first order in $g^2 N$ \cite{Str}. The ultimate challenge is to solve
the closed, linear loop equation for the SU(2) WLA in four dimensions and
to generalize the solution to the case of the SU(3) WLA to solve the
riddle of confinement in QCD.

\vspace{10 mm}
\begin{tabbing}
{\Large\bf Acknowledgements}
\end{tabbing}

I would like to thank
A. A. Migdal for many helpful discussions, C. Callan for a useful comment
and K. Rajagopal and S. Vandermark for careful reading of the
manuscript. This work was partly
supported by the Finnish Cultural
Foundation and the Emil Aaltonen Foundation.

\bibliographystyle{plain}

\pagebreak
\begin{tabbing}
{\Large\bf Figure Captions} \\
\\
Fig. 1. \=
Graphical representation of the non-nested WLA $W_1(\C) =
W_1(\C_{xy}\C_{yx})$ \\
\> $= W_1(S_a,S_b)$ where the WLA is represented by a loop equivalent to
$\C$ \\
\> (having the same topology and areas of the windows as $\C$) and where \\
\> the path $\C_{xy}$ ($\C_{yx}$) with $x=y$ is the boundary of the area $S_a$
($S_b$). \\
\\
Fig. 2. \>
(a) A self-intersection and \\
\> (b) a broken self-intersection at a point $x$. \\
\\
Fig. 3. \>
Graphical representation of a nested WLA $W_{3}^{nested}$. \\
\\
Fig. 4. \>
Graphical representation of the WLA $W(S_1,S_{21},S_{22},S_3)$. \\
\\
Fig. 5. \>
Graphical representation of \\
\> (a) a WLA $W(\C_1,\C_2) = \frac{1}{2} [W(\C_1 \tilde{\C}_2 )
+W(C_1 \tilde{C}_2^{-1} ) ]$ with a composite \\
\> $\;\;\;\;\,$ contour, \\
\> (b) a WLA $W(\C_1, \tilde{\C}_2 ) = W(\C_1, \C_2)$ with a self-touching
contour, \\
\> (c) a WLA $W(C_1 \tilde{\C}_2 )$ with a
self-crossing contour, and \\
\> (d) a WLA $W(\C_1 \tilde{\C}_2^{-1} )$
with a simple loop. \\
\\
Fig. 6. \>
An example of the calculation of the indices $Q$ and $R$. $L_{QR}$ is a \\
\> directed open path from the external infinite area $S_{00}$ to the window \\
\> $S_{QR}$ where $Q$ ($R$) is the number of times the contour $\C_1 \C_2$
crosses \\
\> the path $L_{QR}$ from left to right (from right to left). \\
\\
Fig. 7. \>
Graphical representation of \\
\> (a) the WLA
$W(S_{10},S_{20},...,S_{60},S_{701},S_{702},S_{703},S_{704},S_{80},S_{90})$
and \\
\> (b) the WLA $W(S_{10},S_{20},...S_{60},S_{702} +S_{704},S_{703})$ \\
\\
\end{tabbing}

\end{document}